\documentclass{article}
\usepackage{spconf,amsmath,graphicx, url}
\usepackage[T1]{fontenc}
\def\vec{\mathbf}

\title{A Ray Launching Approach for Computing Exact Paths with Point Clouds}
\name{Niklas Vaara \qquad Pekka Sangi \qquad Miguel Bordallo L\'opez \qquad Janne Heikkil\"a\thanks{This work was supported by the Research Council of Finland (former Academy of Finland) 6G Flagship Programme (Grant Number: 346208), and the Horizon Europe CONVERGE project (Grant 101094831). \\\\ \copyright 2024 IEEE. Personal use of this material is permitted. Permission from IEEE must be obtained for all other uses, in any current or future media, including reprinting/republishing this material for advertising or promotional purposes, creating new collective works, for resale or redistribution to servers or lists, or reuse of any copyrighted component of this work in other works.}}
\address{Center for Machine Vision and Signal Analysis\\ University of Oulu, Finland}
\begin{document}
%
\maketitle

\begin{abstract}

Ray tracing is a deterministic method that produces propagation paths between a transmitter and a receiver. The simulation accuracy is significantly influenced by the environment details. One way to capture the environment with great precision is the utilization of depth sensors and cameras. Such reconstructed environment is in the form of a point cloud. However, utilizing such data directly in ray tracing, a key aspect on vision-aided wireless communications, often involves a significant trade-off between accuracy and execution time. In this paper, we propose an open source novel and fast point cloud-based ray launching algorithm that produces exact paths, 
which provide a good basis for accurate modeling of radio channel
characteristics. 
In experiments, preliminary validation of the ray tracer output is obtained with the aid of a commercial ray tracer.
\end{abstract}


\section{Introduction}
The need for deterministic radio channel characterization methods such as ray tracing (RT) methods  is increasing as new 6G research topics in communications are investigated. For example, development of integrated sensing and communications (ISAC) applications require physically accurate simulations that cannot be achieved with stochastic channel models \cite{hoydis2022sionna}.
Also exploitation of machine learning (ML) methods
requires realistic simulations as it may be hard to get sufficient data from measurement campaigns. 

One cornerstone for accurate RT
simulations is having a model
close to reality in terms of 
spatial details and material properties.
This is envisioned to be achieved by the fusion of multi-modal data acquired from sensors, for example, cameras and depth sensors such as LIDARs, and RADARs \cite{alkhateeb2023real} and is pivotal in achieving high precision in RT simulations. Multi-modal data is beneficial for material segmentation as well, which has been shown by the results in recent works \cite{liang2022multimodal, reza2023multimodal}.

The constructed model acquired from sensor data is in the form of a point cloud. RT-based methods often utilize triangle-mesh based models due to the clearly defined surfaces and wide adaptation in the field of computer graphics. However, point clouds have also been utilized for radio channel characterization in varying settings \cite{koivumaki2023ray, jarvelainen2016indoor, koivumaki2021impacts}, and their use in this context represents a significant step towards the integration of radio and vision data for enhanced environmental modeling.

\begin{figure}[t]
    \centering
    \includegraphics[width=1.0\linewidth]{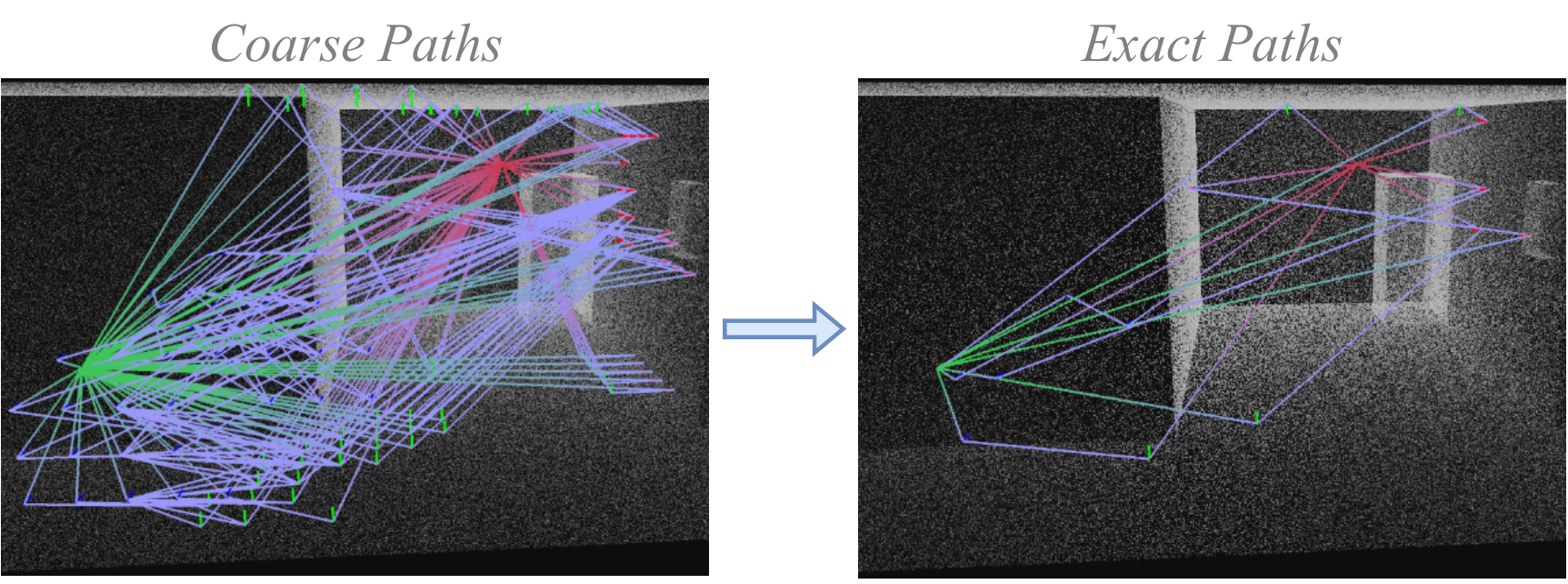}
    \caption{Coarse paths with two or fewer reflections and the remaining exacts paths generated with our implementation.}
    \vspace{-3mm}
    \label{fig:c2r}
\end{figure}

RT-based algorithms can be categorized into image-based and ray launching (RL) algorithms \cite{yun2015ray}. The former method focuses on finding the exact paths for a transmitter/receiver (TX/RX) pair in a given environment, while in RL rays are cast from the TX with uniform spacing without considering the RX locations. RL has generally higher performance, but degraded accuracy due to discrete ray directions and approximated ray reception at RXs. Most RT solutions that directly utilize point clouds adopt the image method. Inspired by \cite{lu2018discrete}, an environment-driven RL-based solution was introduced in \cite{koivumaki2022point}, which utilizes downsampled point clouds to compute coarse approximations of the propagation paths. As a pre-processing step, a matrix is filled that contains information about the visibilities between the points, TXs, and RXs in the environment. This matrix generation process becomes highly time consuming with a large number of points. 

In this paper, we present an open source\footnote{https://github.com/nvaara/NimbusRT} novel GPU-based RL solution that directly utilizes point clouds in the path determination process. 
While requiring minimal pre-processing measured in terms of execution time, it provides an approach that allows fast adaptability particularly in dense and dynamically changing environments. As illustrated in Fig.~\ref{fig:c2r}, our approach first produces coarse paths consisting of reflections and diffractions with a method, which utilizes multi-level voxel-based representation of the environment. As a post-processing step, we propose a dynamic solution to refine the coarse paths to exact paths.

\vspace{-3mm}
\section{Description of the Simulator}

The simulator consists of three GPU-based components: (1) voxelization, (2) coarse path tracing, and (3) path refinement. It produces exact paths in labeled point cloud environments between a number of TXs and RXs. Labeled point cloud refers to a point cloud, where each point contains at least a position, normal vector, and a label. The labels are acquired by a segmentation process, where the resulting clusters represent distinct surfaces. 
It also supports diffraction, although the diffraction edges have to be inserted manually.

\vspace{-3mm}
\subsection{Voxelization}
\label{ss_voxelization}

Point clouds lack the explicit surface representation of triangle mesh-based models, which is why we discretize the environment into three dimensional small uniform volumetric axis-aligned elements known as voxels. Such structure enables cheap spatial computations. The volume of each voxel represents a part of the original geometry, which allows us to examine a small area inside a scene. This is beneficial especially with point clouds, where the geometry may be represented with densely located points.

We utilize both low and high resolution voxels. The low resolution voxels are represented in the form of a grid, where traversal operations are performed. Having a low  
resolution greatly reduces the memory usage  
and allows covering
a broader volume during traversal operations, which is beneficial as we utilize conical rays. On the other hand, high resolution allows us to have more detailed localization of intersections.

\begin{figure}[htbp]
    \centering
    \includegraphics[width=0.85\linewidth]{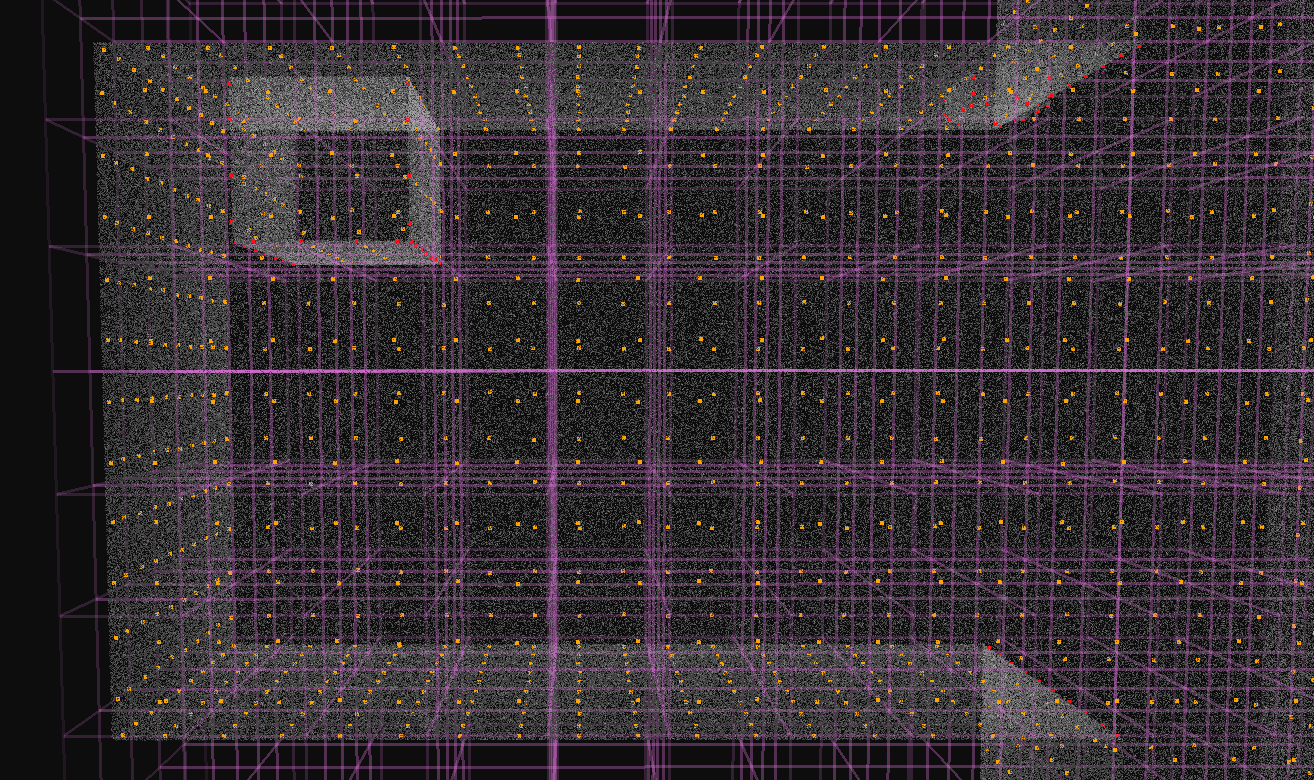}
    \caption{A point cloud consisting of white points, purple low resolution voxel grid, and ray reception points of intersectable entities as orange and red points.}
    \vspace{-3mm}
    \label{fig:voxelization}
\end{figure}

Each high resolution voxel, which we refer to as a subvoxel, is obtained by spatially dividing a low resolution voxel with a factor $D_v$. The subvoxels are only used for forming discretized parts of the geometry bound by its faces, and each of them may contain diffraction edge segments, points from the point cloud, and RXs. We call these alternatives as intersectable entities (IEs). On top of the geometry, each IE contains a label and a ray reception point. Each low resolution voxel is aware of its IEs, which means that they do not need to be stored in memory consuming data structures. A voxel grid and IEs are illustrated in Fig. \ref{fig:voxelization}.

\subsection{Coarse Path Tracing}

The coarse path tracing outputs approximate propagation path trajectories. The path determination consists of two distinct phases, transmission and propagation, which are performed for each TX separately.

\subsubsection{Intersections}

Intersections with the point cloud have to be approximated, as they do not have a concise surface representation like triangles. We leverage the following procedure to determine intersections between a ray and a point set. The intersection point is determined with a signed distance function (SDF). We use the following implicit surface equation to estimate the distance to the surface \cite{adamson2004approximating}:
\begin{equation}
    \label{fSdf}
    f_{sdf}(\vec{x}) = \left(\vec{x} - \vec{p} \left(\vec{x}\right) \right) \cdot \vec{n}\left(\vec{x} \right), 
\end{equation}
where
$\vec{x}$ is a position being evaluated for intersection, $\vec{p}(\vec{x})$ and $\vec{n}(\vec{x})$ represent the weighted averages of the nearby point positions and normals, respectively. We compute a point on the ray by orthogonally projecting the center point of a bounding box enclosing the point set to the ray. This point acts as the center point for a line segment. The line segment length is the diameter of the subvoxel. Starting from the line segment end point $\vec{s}_0$ closest to the ray origin, we advance along the ray direction vector with the distance given by $\left| f_{sdf} \left(\vec{s}_i \right)\right|$ until an intersection is found or the end of the line segment is reached. An intersection is found if $ \text{sign} \left( f_{sdf} \left(\vec{s}_i \right)\right) \neq  \text{sign} \left( f_{sdf} \left( \vec{s}_{i+1} \right) \right)$ or if $\left| f_{sdf} \left( \vec{s}_{i} \right) \right|$ is near zero.

\subsubsection{Transmission Phase}

The path tracing process starts with the transmission phase. We determine the initial interaction points from each TX with a similar method present in environment-driven approaches \cite{lu2018discrete, koivumaki2022point}, where rays are only launched towards discretized parts of the geometry. In our approach, a ray is cast towards each IE present in the scene, resulting in line of sight (LOS) connections to RXs and initial IE interactions that are visible from the TX.

\subsubsection{Propagation Phase}

After the transmission phase, new rays are launched from the initial IE interaction points. Depending on the interaction point type, the rays are either reflected or diffracted. The rays are traced until they do not intersect with any IEs or if maximum number of interactions is reached. As the rays formed in propagation phase are not cast towards IEs, the voxel grid presented in Section \ref{ss_voxelization} is utilized for traversal. Together with conical rays (see Fig. \ref{fig:rays}), we refer this method of traversal as voxel cone tracing. Due to the conical rays, multiple similar paths are generated during the path tracing process. We define $\kappa$ as the maximum number of paths that can contain the same labels in the same order.

\vspace{-3mm}
\subsubsection{Voxel Cone Tracing}

In the traversal we use ray marching \cite{pharr2005gpu}, which refers to advancing a varying distance during each iteration. The march distance from a given voxel is the longest distance along any axis to the closest voxel that contains IEs. Each voxel includes the information whether it contains IEs, as well as the march distance to the closest voxel containing IEs. If the voxel being evaluated contains IEs or if the march distance is equal to one, all of the neighboring voxels are evaluated for intersections.

We also utilize conical rays similarly to \cite{koivumaki2022point}. This is beneficial especially for diffractions, that are often simulated by casting rays in the form of a Keller's cone \cite{keller1962geometrical}. The diffracted rays experience increasing spatial separation with the neighboring diffracted rays, which in traditional methods is often reduced by ray splitting \cite{fortune1998efficient} with additional computational cost. This can be avoided with conical rays as the rate of increase in the spatial separation between the neighboring diffracted rays can be expressed with an angle. 

When a voxel is being evaluated for intersections, we form a bounding sphere around it for cone-sphere intersection. If the intersection exists, a similar intersection test is performed for each IE inside the voxel using a bounding sphere that encompasses IE's subvoxel. For each valid cone-sphere intersection with an IE, the visibility is validated by tracing a ray from the ray origin to the IE reception point.

\begin{figure}[htbp]
    \centering
    \includegraphics[width=1.0\linewidth]{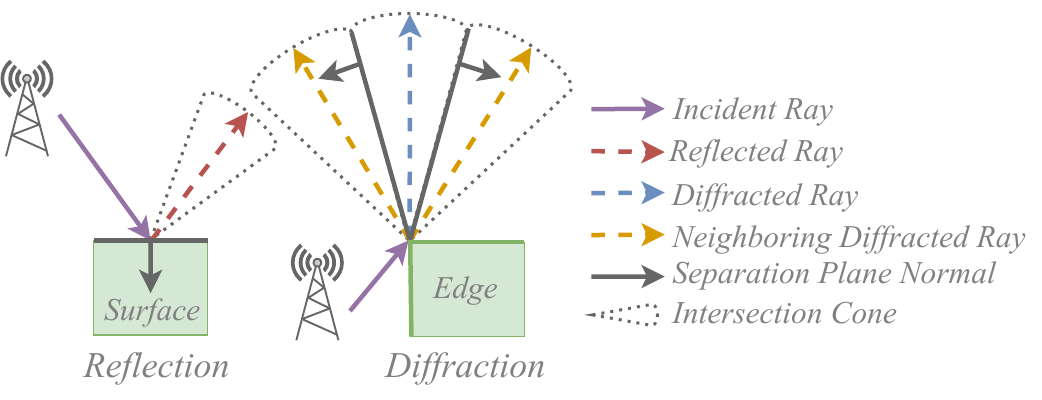}
    \caption{Formation of intersection cone and separation plane normals for a reflected and a diffracted ray.}
    \vspace{-3mm}
    \label{fig:rays}
\end{figure}

The cone apex angle is constant for all of the conical rays. In the case of diffractions, we mitigate the duplicate paths by introducing two separation planes. These planes, located in between the two neighboring diffracted rays, are used to compute the signed distance to the IE reception point, which we utilize to determine whether the IE reception point lies between the planes. In the case of reflected rays, a separation plane is placed at the origin of the ray. The components of the aforementioned rays are illustrated in Fig.~\ref{fig:rays}.

\vspace{-3mm}
\subsection{Path Refinement}

The resulting paths produced by coarse path tracing are not accurate due to the discretized interaction points and conical rays. As a post-processing step, the image method is often applied to refine the coarse paths into exact paths, as in \cite{hoydis2023sionna, choi2023withray}. However, refining diffraction interaction points is not directly possible with the image method and requires iterative methods.

We build our path refiner on top of our previous work \cite{vaara2023refined}, where the exact paths are computed for reflections and diffractions by minimizing the path length. The previous work assumes infinite boundless planes for reflection points, which we address in this work. The mathematical foundation is the same, which we describe briefly. The path length to be minimized can be described as
\begin{equation}
    f = \sum_{k=0}^{N+1} ||\vec{I}_{k} - \vec{I}_{k+1}||,
\end{equation}
where $\vec{I}_{k}$ is the $k$th interaction point. $\vec{I}_{0}$ and $\vec{I}_{N+1}$ are the TX and RX locations, respectively. The local path length of each interaction point is
$ \label{f_k} f_{k} = ||\vec{I}_{k} - \vec{I}_{k - 1}|| + ||\vec{I}_{k} - \vec{I}_{k + 1}||\text{.}$
In the case of an $k$th intersection point being a reflection, the reflection point is
$\vec{I}_{k} = \vec{R}_{k} = \vec{c}_{k} + r_{k} \vec{u}_{k} + s_{k} \vec{v}_{k},$    
where $\vec{c}_k$ is the current interaction position, $\vec{u}_{k}$ and $\vec{v}_{k}$ are the vectors that form an orthonormal basis with the surface normal vector at $\vec{c}_{k}$, and lastly, $r_k$ and $s_k$ are the variables to solve. Similarly, the $k$th interaction in the case of a diffraction is $\vec{I}_{k} = \vec{D}_{k} = \vec{c}_{k} + t_{k} \vec{w}_{k},$ where $\vec{c}_k$ is the current interaction position, $\Vec{w}_{k}$ is the direction along the edge, and $t_k$ is the variable to solve.

During each iteration, we compute the partial derivatives $\partial f_{k} / \partial r_{k}$ and $\partial f_{k} / \partial s_{k}$ for each reflection point, and $\partial f_{k} / \partial t_{k}$ for each diffraction point. The interaction points are then updated using gradient descent. The reflection points require additional processing, as they are acquired from the point cloud, which we do not impose any restrictions to in terms of the geometry. Thus, we trace a ray in the direction of the updated point to determine the new $\vec{c}_{k}$, $\vec{u}_{k}$ and $\vec{v}_{k}$. This process is performed for $\rho$ iterations, or until a ray misses. If $\rho$ iterations are reached, the path is considered valid if the norm of gradient vector is less than the convergence threshold $\delta$. Lastly, we validate the visibility between each node on the path trajectory by RT.

Label-based duplicate path removal is applied to refined paths as well, however, we only keep the shortest path with a certain label and interaction combination. An additional post-processing step is beneficial, as all of the duplicate paths may not be recognized due to reflections near edges that result in different labels. First, we sort the remaining paths based on the time delay of the path. Similarly to \cite{jarvelainen2016indoor, koivumaki2022point}, we compute the first Fresnel zones to validate the uniqueness of each path. Among the paths that contain the same chain of interactions and where the interaction points lie within the Fresnel zones, the shortest path is kept. Recall Fig.~\ref{fig:c2r} for an example of the coarse paths and remaining exact paths.
\vspace{-2mm}
\section{Experiments}

In the experiments, we evaluate the performance of our simulator, validating the generated paths by comparing our results with baseline paths acquired with a commercial ray tracer.

\vspace{-3mm}
\subsection{Overview}

\begin{figure}[htbp]
    \centering
    \includegraphics[width=1.0\linewidth]{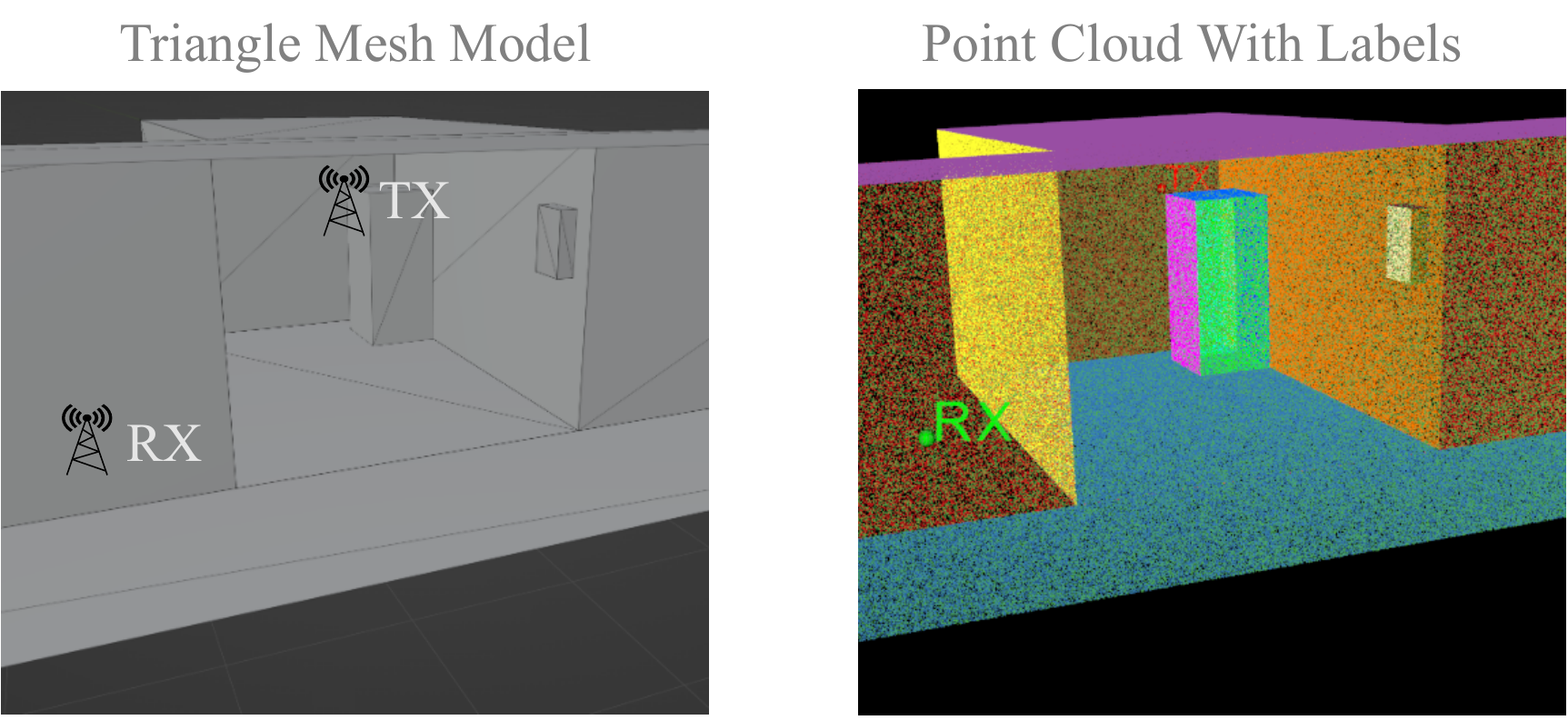}
    \caption{Triangle mesh and point cloud version of the corridor environment. One wall missing for visualization purposes.}
    \vspace{-5mm}
    \label{fig:models}
\end{figure}

The experiments are performed with a synthetic point cloud generated from a triangle mesh version of a corridor model, a typical indoor environment. The resulting point cloud contains 2745513 points. Both models are presented in Fig. \ref{fig:models}. The labels for the point cloud were generated with a plane fitting algorithm available in PCL \cite{rusu20113d}. As the baseline, we utilize Wireless Insite \cite{wirelessinsite}, which is a commercial RT tool that utilizes triangle-mesh based models and has been validated by measurements, for example, in \cite{suga2023indoor, mededjovic2012wireless}. 
Paths for comparison were generated
with its X3D RT model.

\subsection{Simulations}
The desktop computer used in the simulations contains the following components: AMD Ryzen 9 5900X CPU, NVIDIA GeForce RTX 3080 GPU, and 32GB of RAM. Two non-LOS simulations were performed with one TX and one RX. The used simulation parameters, reflecting typical conditions in ISAC scenarios, are presented in Table \ref{tab:params}. The antennas are isotropic and their locations can be seen in Fig. \ref{fig:models}. Simulation A was performed without diffractions and simulation B was conducted with a diffraction limit of one. In the case of diffractions, we only consider exterior edge diffractions. The baseline paths were generated with the triangle mesh version. Path matching between the baseline and our simulator was performed by comparing the number of interactions and their types, and ray trajectories. If the difference in the ray trajectories of the paths being compared was less than one degree and the number of interactions and types were identical, the path was considered to be a match. The results with execution times can be found in Table \ref{tab:results}.

\begin{table}[t]
    \centering
    \caption{Simulation parameters.}
    \begin{tabular}{cc}
    \hline
         Carrier Frequency (GHz) & 60 \\ 
         Voxel Size (m$^3$) & 0.5\\
         Voxel Division Factor $D_v$ & 2 \\
         Coarse Path Duplicate Limit $\kappa$ & 100 \\
         Maximum Number of Interactions & 5 \\
         Convergence Threshold $\delta$ & 0.0001 \\
         Refinement Iterations $\rho$ & 2000\\
         \hline
    \end{tabular}
    \vspace{-3mm}
    \label{tab:params}
\end{table}

\begin{table}[t]
    \centering
    \caption{Simulation results. Simulations A and B were conducted with a diffraction limit of zero and one, respectively.}
    \begin{tabular}{ccc}
        \hline
         Simulation & A & B \\
         \hline
         Coarse Path Tracing (s) & 7.65 & 150.02\\
         Path Refinement (s) & 2.80 & 26.35 \\
         Coarse Paths & 105730 & 755136\\
         Exact Paths & 90 & 1480\\
         Baseline Paths & 83 & 654\\
         \% of Baseline Paths Found & 100 & 85.8\\
         \hline
    \end{tabular}
    \vspace{-3mm}
    \label{tab:results}
\end{table}

\vspace{-3mm}
\subsection{Discussion}

The experiments show that the simulator can produce similar paths as the baseline. Some of the paths found by our simulator that contain a reflection may not be accurate representations of the exact path due the interaction point being located near an edge, which causes distortion in the approximated normal vector.

In the case of the paths containing a diffraction, our simulator did not find all the baseline paths. One reason for this is that rays are only cast towards the IE reception points. This may lead to some paths being missed, especially if the following interaction point is near. An example of such is the box on the wall in Fig. \ref{fig:models}. One way to mitigate this issue is to utilize a smaller voxel size or a larger voxel division factor $D_v$ with the cost of increased computational complexity.

To make comparison with the reference method, these experiments considered perfect synthetic point clouds. However, the implementation can be used with raw point clouds acquired directly from the sensors, which will be one topic in future work. In addition, comparisons against channel measurements need to be conducted.

\vspace{-3mm}
\section{Conclusion}

We presented an open source novel ray launching algorithm that utilizes point clouds in the path determination process. The method first generates coarse path approximations, which are then corrected to exact paths in the path refinement phase. The results of preliminary validation show substantial similarity with the baseline paths generated with a commercial ray tracer. 

\bibliographystyle{IEEEbib}
\bibliography{citations}

\end{document}